\documentclass[12pt,technote]{IEEEtran}

\usepackage{ifpdf}
\usepackage{graphicx}
\usepackage[space]{grffile}
\usepackage{latexsym}
\usepackage{textcomp}
\usepackage{longtable}
\usepackage{multirow,booktabs}
\usepackage{amsfonts,amsmath,amssymb}
\usepackage{url}
\usepackage{hyperref}
\hypersetup{colorlinks=false,pdfborder={0 0 0}}

\newif\iflatexml\latexmlfalse
\usepackage[utf8]{inputenc}
\usepackage[english]{babel}

\usepackage{cite}

\usepackage{array}
\usepackage[caption=false,font=footnotesize]{subfig}

\usepackage{fixltx2e}

\usepackage{url}

\DeclareMathOperator*{\argmax}{argmax}

\newcommand{\unit}[1]{\ensuremath{\mathrm{\,#1}}}

\begin{document}
%

\title{Body movement to sound interface with vector autoregressive hierarchical hidden Markov models}

\author{Dimitrije Markovi\'c, Borjana Val\v{c}i\'c, and Neboj\v{s}a Male\v{s}evi\'c
\thanks{Dimitrije Markovi\'c is with Chair for Neuroimaging, Department of Psychology, Technische Universit\"at Dresden, 
01069 Dresden, Germany (email: dimitrije.markovic@tu-dresden.de).}
\thanks{Borjana Val\v{c}i\'c is with Faculty of Music, University of Belgrade, 11000 Belgrade, Serbia (email: valcicborjana@gmail.com)}
\thanks{Neboj\v{s}a Male\v{s}evi\'c is with Faculty of Electrical Engineering, University of Belgrade, 11000 Belgrade, Serbia, 
and Dept. of Biomedical Engineering, Lund University, SE-221 00 Lund, Sweden (email: nebojsa.malesevic@etf.rs)}
}

\maketitle


\begin{abstract}
Interfacing a kinetic action of a person to an action of a machine system is an important research topic in many application areas.  One of the key factors for intimate human-machine interaction is the ability of the control algorithm to detect and classify different user commands with shortest possible latency, thus making a highly correlated link between cause and effect. In our research, we focused on the task of mapping user kinematic actions into sound samples. The presented methodology relies on the wireless sensor nodes equipped with inertial measurement units and the real-time algorithm dedicated for early detection and classification of a variety of movements/gestures performed by a user. The core algorithm is based on the approximate Bayesian inference of Vector Autoregressive Hierarchical Hidden Markov Models (VAR-HHMM), where models database is derived from the set of motion gestures. The performance of the algorithm was compared with an online version of the K-nearest neighbours (KNN) algorithm, where we used offline expert based classification as the benchmark. In almost all of the evaluation metrics (e.g. confusion matrix, recall and precision scores) the VAR-HHMM algorithm outperformed KNN. Furthermore, the VAR-HHMM algorithm, in some cases, achieved faster movement onset detection compared with the offline standard. The proposed concept, although envisioned for movement-to-sound application, could be implemented in other human-machine interfaces.
\end{abstract}%

\IEEEpeerreviewmaketitle


\section{Introduction} 
\IEEEPARstart{T}{raditional} musical instruments generate sounds as a result of interaction between a targeted musician's movement and the inherent physical properties of the instruments. However, when playing music an experienced musician does not move purely in a way that allows him to perform optimally on a particular instrument, but in a way that enables him to communicate personal experiences to an audience \cite{jensenius2007action,gritten2006music,cadoz2000gesture}. With a rapid progress of digital technology, it became possible to separate human movement from the direct sound formation \cite{bongers2000physical,paradiso1997electronic}. Although expanding possibilities of an artist to express itself through the custom sound waveforms, this leap has diminished the necessity for performer's body movements; thus it has reduced the capacity of music and social interaction between a performer and an audience \cite{ferreira2008sound,stuart2003object,schloss2003using}. 

One of the novel tendencies among digital instruments designers and artists, in the digital music act, is to increase somatic and corporeal presence  \cite{tanaka2000musical} and induce a rousing connection between movement and sound production. These modern music performances comprise of exaggerated gestures that are suitable for interfacing a digital controller \cite{winkler1995making}. One of the most important requirements for creating an emphatic connection between a performer and an audience is a clear and coherent process of movement based sound production \cite{mitchell2011soundgrasp,wang2008chuck,bevilacqua2007wireless,iazzetta2000meaning,choi1998motion,goldstein1998gestural}. \cite{bahn2001physicality} have emphasised that this principle should be met regardless of the instrument's construction, incorporated technology, and preferred playing style. Although the number of contemporary (body movement orientated) digital instruments and scientific publications related to them is rapidly increasing, there is a lack of systematic approaches for defining the methods used to detect and classify human movements designated as digital instrument input.

In this work, we have designed a digital instrument that is driven by body movements. The instrument comprises of several sensors (accelerometers and gyroscopes) integrated into small wireless units positioned on various body parts. The system is executing a real-time movement classification, using individual movement trajectories, where each gesture corresponds to a sound sample from the audio database. We named this digital music instrument "Movezik". We coupled the creation of this performance-oriented digital instruments with the development of novel human-machine interaction algorithms capable of establishing a causal link between human action and machine generated effect. We have formulated the classification algorithm as an online Bayesian classifier, in which we use a hierarchical hidden Markov vector autoregressive process (VAR-HHMM) \cite{ephraim2002hidden,yang2000some} to model dynamics of the real-time recordings of movement trajectories. We estimated the free parameters of the dynamical models---which approximate the movement dynamics---using a combination of two often used parameter estimation algorithms, the Expectation Maximization (EM) \cite{north1998learning} and the Viterbi algorithm \cite{logothetis1999expectation,forney1973viterbi}. Importantly, we compared the performance of VAR-HHMM with the k-nearest-neighbour (KNN) classifier, which showed highest classification accuracy, on the recorded movement data, among several classification algorithms, such as neural network classifier \cite{bishop1995neural}, quadratic and linear discriminant classifiers \cite{cacoullos2014discriminant,lachenbruch1979discriminant}, and support vector machines \cite{meyer2015support,du2014support}.

We quantified the classification performance of the two algorithms (KNN and VAR-HHMM) using well-established classification metrics \cite{davis2006relationship,fawcett2006introduction}, such as confusion matrix, precision score (PS), and recall score (RS). Also, to qualitatively evaluate the algorithms' performance, we have designed and conducted a behavioural experiment. In the experiment, the human participants were asked to rate synchronisation levels of a pre-recorded movement and a short sound that was delayed (for a randomised amount of time) relative to the movement onset. We used the collected ratings to derive confidence interval of acceptable latencies, which we have set as the upper bounds on the algorithms' performance.

Besides the direct implementation of this setup to a movement-sound mapping, the classification method presented here can also be applied to general human-machine interfaces that require a variety of highly synchronised movement triggered commands \cite{haptic}. Specifically, the core algorithm can be implemented as the link between continuous actuators (human commands) and an execution of discrete robotic actions, e.g. in haptic robots, exoskeletons, artificial limbs, and wheelchairs.

\section{Methods}
\subsection{Hardware system and movement recordings}
The Movezik's hardware system comprises of up to five sensorized wireless nodes connected to the multi-platform master application. Each node is dedicated to acquiring data from inertial measurement unit (IMU) with three axial accelerometer and three axial gyroscope. The IMU's data was sampled at fixed frequency of $100 \unit{Hz}$ and amplitude resolution of $16$ bits, whereas the sensitivity of the accelerometers and the gyroscopes was fixed to a predefined range. We have set the accelerometer's range to $[-2\unit{g}, 2\unit{g}]$---where $\unit{g}$ denotes the value of the nominal gravitational acceleration---and the gyroscope's range to $[-500\unit{\frac{^{\circ}}{s}}, 500\unit{\frac{^{\circ}}{s}}]$. By connecting this wireless inertial sensor device (WISD) to an arbitrary body part (e.g. hands, legs, head, {\it etc.}) we can map the movements of that body part into the 6D state space defined by the IMU sensors. 

\begin{figure}[!ht]
\centering
\includegraphics[width=\columnwidth]{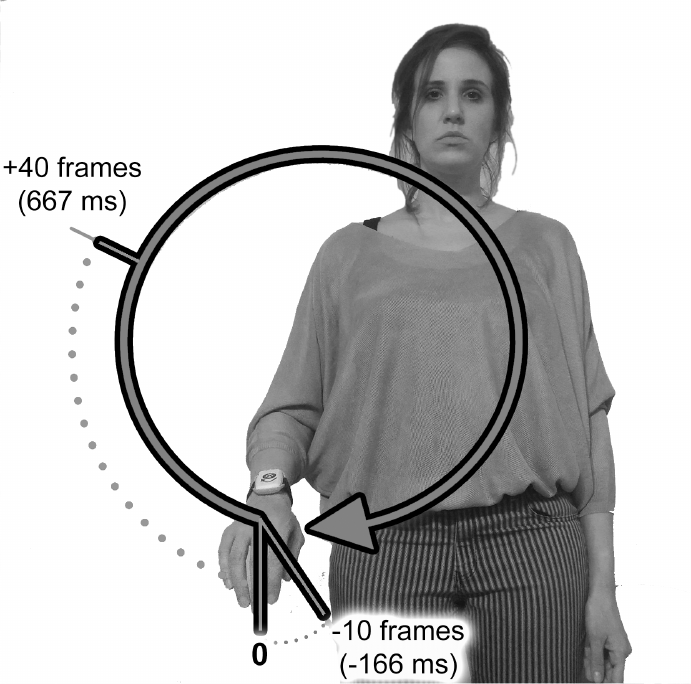}
\caption{Subjective movement-to-sound criterion (SMtSC) measurement setup. In each trial, sound sample is played in a pseudo-random manner in the range $[-10, 40]$ frames with respect to the onset of the clockwise circular hand movement. The wireless IMU node of the Movezik system is positioned on the wrist of the right hand.}
\label{fig:experiment}
\end{figure}

We have recorded in total four different data sets using a WISD attached to the right hand (see Tab.~\ref{tab:summary} for details). The three out of four data sets consist of two complementary hand movements separated by resting periods (a stable hand position). The fourth data set consists of intermittent circular hand movements. An example of such circular hand movement and attached WISD is shown in Fig.~\ref{fig:experiment}.

\subsection{Subjective movement-to-sound criterion (SMtSC) table}
The amount of acceptable latency between human kinetic action and associated sound is the main evaluation criterion for the performance of the proposed VAR-HHMM algorithm. Understanding and quantifying human perception of the movement-sound correlation is the interesting topic, which could also be useful for multimedia artists and engineers involved in similar projects. For the purpose of quantifying human perception we have designed a behavioral experiment in which we recorded participants estimates of a causal link between a movement and a lagged sound \cite{vatakis2006audiovisual, zampini2005audio}.

To exclude participants relating sound to arbitrary segments of the movement (classifying the end or an intermediate state of the movement as synchronized with sound onset) we used as a visual stimulus a recording of a continuous and smooth circular arm movement that was paired with an abbreviated sound sample. The duration of the clockwise circular arm movement was around $2.5\unit{s}$ and was preceded and succeeded by shorter periods of resting hand position (see Fig. ~\ref{fig:experiment}). The video of the circular movement was recorded with a high-speed camera at $120\unit{fps}$, and later down-sampled to $60\unit{fps}$ for displaying on regular monitors. The total duration of the recording was set to be $3.5\unit{s}$ ($0.5\unit{s}$ of steady periods before and after the $2.5\unit{s}$ long circular movement). The sound sample was synthesized from a single digital piano tone ($a1$) and additionally shortened by multiplying the recorded sound with the $\frac{1}{t}$ function, resulting in the short sound that has more than $90\%$ of its power within $20\unit{ms}$. The exact moment of the movement onset was identified by visual inspection of the high-speed video. The sound latency marker (delay between the movement onset and the sound onset) was defined using the visually determined movement onset marker. 

To automate the experimental procedure we have developed a python script based on PsychoPy application \cite{peirce2007psychopy} that generated sample videos (trials) with pseudorandom sound latencies ranging from $-10$ to $+40$ frames ($-166\unit{ms}$ to $667\unit{ms}$), looping the trials until subject's decision and logging response and decision time per each trial. The order of trials was the same for each subject, predefined as a random combination of trials ($51$ of them) repeated three times (in total $153$ trials per subject). The experimental design with three repetitions of complete trials sets allowed us to: (i) excluded participants with high variability of responses to the same trial (not consistent participants); (ii) estimate the latency that defines a boundary between trials classified as synchronous and asynchronous.

Movement-to-sound synchronization was estimated based on responses from $20$ participants of different age ($31\pm 12$ years). During the measurement protocol we used laptop PC with LCD monitor (size $15.6$ inch and resolution $1920$x$1080$) positioned at eye level, approximately 1 m away from the subject. Subjects were receiving  sound sample through in-ear headphones at self-preferred volume and were instructed to press the left arrow key to rate a trial as synchronous and the right arrow key to rate a trial as asynchronous.

\subsection{Offline classification algorithm}
\begin{figure}[!ht]
\centering
\includegraphics[width=\columnwidth]{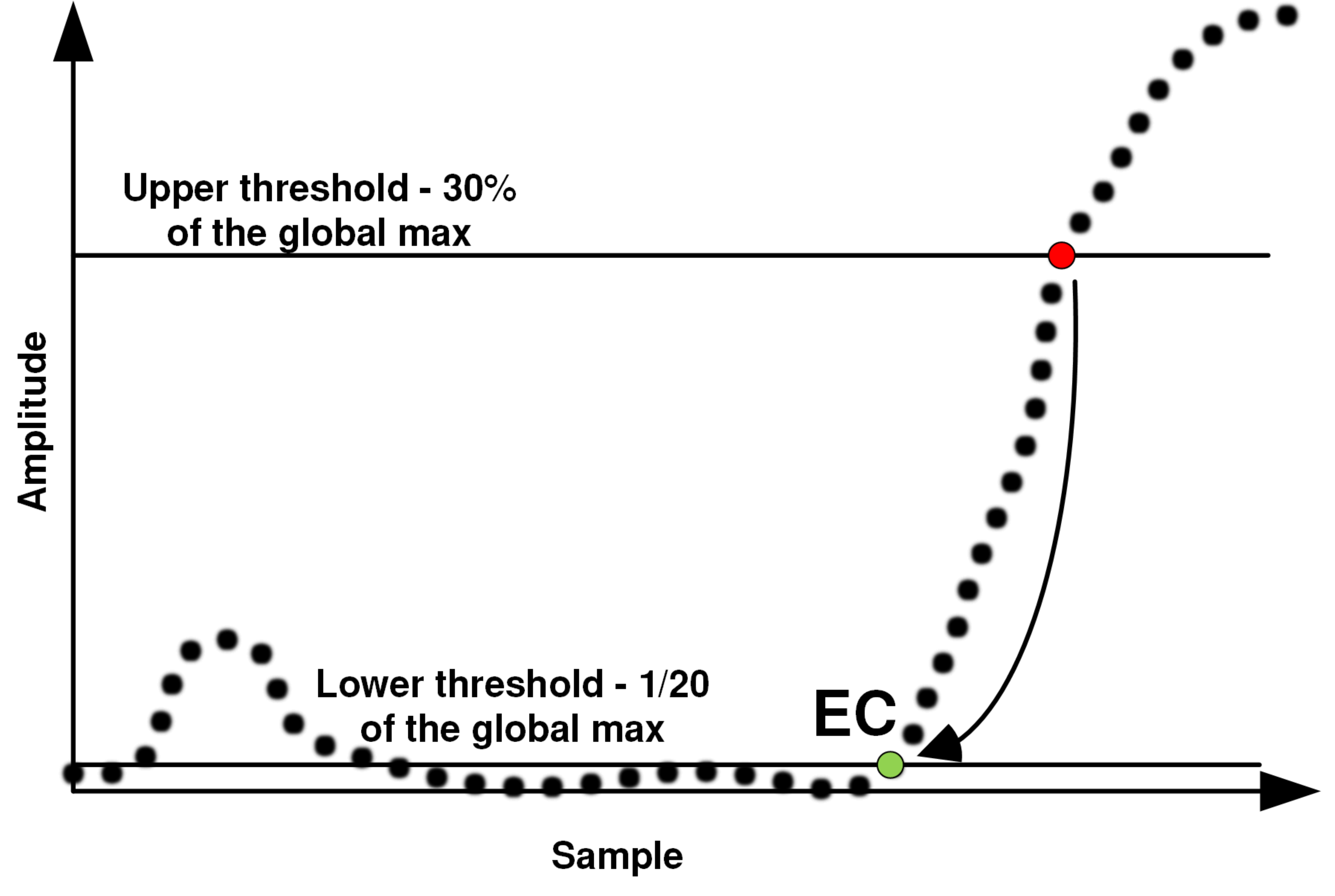}
\caption{Illustration of the offline classification algorithm with the two threshold method. Algorithm moves sample by sample until the upper threshold is exceeded (red dot); then, it cycles back until the value of the signal goes below of the lower threshold. The time point related to this value (green dot) is set as the movement onset.}
\label{fig:EC}
\end{figure}

The expert classification (EC) method is an offline, threshold based algorithm combined with a checkup from the human expert. The threshold based algorithm follows the conventional two threshold methods (Fig. ~\ref{fig:EC}). It consists of an upper threshold for reliable movement detection and a lower threshold for detection of the movement onset. The algorithm goes as following: when a signal exceeds the upper threshold the algorithm searches previous samples until the lower threshold is reached; 
the sample on which the lower threshold is reached is stored as axis movement onset marker. Similarly, axis movement ending is marked at the point after the upper threshold exceeding where a signal reaches the lower threshold after. The algorithm for movement detection uses three parallel loops for three axes of rectified gyroscope signal. The upper threshold is set to $30 \%$ of the maximal gyroscope value while the lower threshold is set to $1/20$ of the upper threshold. For the EC movement onset, first out of three axis movement onset markers from the 3 gyro axes is considered, and for the EC movement end the last out of three axis movement ends is considered. Following the automated algorithm detection of movement, the human expert manually inspects detection and classification to exclude false positives and false negatives.

\subsection{Segmentation of the movement recordings using the expert classifier}
The recorded data sets that we used to test and compare the movement classifiers consisted of four separate data sets. Three out of four data sets combined recordings of two different movements separated by periods in a rest position at the end of the movement. We have recorded the following movements: (i) left-right hand movements ($M_2, M_3$), (ii) up-down hand movements ($M_4, M_5$), (iii) circular hand movements ($M_6$; see Fig.~\ref{fig:experiment}), and (iv) supination-pronation of hand wrist ($M_7, M_8$; elbow was kept in a stable position). The recordings of the hand resting in various position at the end of the movements were labeled as single movement $M_1$. 

The summary of several properties of the four data sets is provided in Table~\ref{tab:summary}. The number of movements in the last column of the table denotes the total number of recorded repetitions of movements within each data set. Hence, for the data sets consisting of two types of movements (e.g. left-right), each movement was repeated for half of the total number of repeated movements. Note that although the data set consisting of circular movement is significantly shorter than the rest, the total number of the recorded data points within the movement is roughly the same as in the other data sets, as the duration of circular movement is longer than that of other movements.

Prior to training the KNN and VAR-HHMM classifiers (see bellow) all the data sets were labeled using the labels provided by the offline classification procedure that is described in previous subsection. The half of the recorded data from each data sets were used for training of the online classifiers and the other half for testing the performance of the two online algorithms.

\begin{table}
\centering
\caption{The summary of the recorded data sets.}
\label{tab:summary}
\begin{tabular}{lll}
\hline
\textbf{Data set}        & \textbf{Number of data points} & \textbf{Number of movements} \\
\hline   
left-right           & 12800      & 20 \\            
up-down       & 15000   & 20 \\
circular        & 3188  &  6 \\
rotational      & 14201  & 20 \\
\hline
\end{tabular}
\end{table}

\subsection{K-nearest neighbors classification algorithm}
Due to its simplicity, easy implementation and widespread use for classification purposes, weighted k-nearest neighbor (KNN) \cite{hechenbichler2004weighted} algorithm was selected as the golden standard, out of the several widespread classifiers, for the time domain signals. To standardize optimization and test procedures we used the scikit-learning python library \cite{scikit-learn}. We fixed the number of neighbours for the KNN classifier to $k = 29$, based on the optimisation procedure that maximised the classification accuracy over the four data sets. 

\subsection{Vector Autoregressive Hierarchica Hidden Markov classifier}
The vector autoregressive Hierarchical Hidden Markov classifier (VAR-HHMM) is based on a probabilistic generative model. The main assumption of the model is that the movement trajectories recorded using the Movezik system can be modeled with a pice-wise linear dynamical system. In other words, we assumed that each movement can be split into segments, where the dynamics of the recorded signal within each segment can be captured by an vector autoregressive (VAR) model. Furthermore, the transition between the segments is stochastic and captured by a Hidden Markov Model (HMM). Similarly, the transition between different movements is also captured by additional Hidden Markov layer that sits on the top of the hierarchy. The half of the recorded and pre-labeled data sets was used for training (fitting the free parameters of the generative model) and the other half for testing the online classification capabilities of the inverted model. The online classification of the recorded signals was performed using an approximate Bayesian inference procedure described bellow, where the label of the current data point corresponds to the movement with highest posterior probability. 

\subsubsection{Generative model}
The general form of an vector autoregressive model is defined as 
\begin{equation}
 \vec{y}_t = \vec{\mu}^{(i,m)} + \sum_{p = 1}^{\tau}A_p^{(i,m)} \vec{y}_{t-p} + \vec{\psi}^{(i,m)}_t
\end{equation}
where $\vec{y}_t$ denotes the recording obtained at time step $t$, and $\vec{\psi}^{(i,m)}_t$ denotes an i.i.d. random normal variable with mean zero. The parameters of the AR($\tau$) model that define the transition from previous to current measurements are denoted with $A_1^{(i,m)}, \ldots, A_{\tau}^{(i,m)}$, the $\vec{\mu}^{(i,m)}$ captures the constant signal value. Importantly, the superscripts $i \in \{1,\ldots,5\}$, and $m \in \{1,\ldots,8\}$ denominate the $i$th segment of the $m$th movement. Note that we have assumed that each movement can be separated on maximally five linear segments.

Thus, we can write the observation likelihood as 
\begin{multline*}
 p\left(\vec{y}_t| \vec{y}_{t-1:t-p}, S_t = i, M_t = m \right) = \\
 \mathcal{N} \left(\vec{y}_t; \vec{\mu}^{(i,m)} + \sum_{p = 1}^{\tau}A_p^{(i,m)} \vec{y}_{t-p}, \Sigma^{(i,m)} \right).
\end{multline*}

The transition matrix of movement segments for the $m$th movement is given as
\begin{equation}
\label{eq:segtm}
\begin{split}
 p\left(S_t = i| S_{t-1} = j, M_{t-1} = M_t = m \right) &= T^{(m)}_{i,j}, \\
 p\left(S_t = i| S_{t-1} = j, M_{t-1} \neq M_t = m \right) &= p_{i,m},
 \end{split}
\end{equation}

where $p_{i,m}$ denotes the prior probability of the $i$th movement segments of the $m$th movement. Similarly, the transition matrix between different movements is given as 
\begin{equation}
\label{eq:movtm}
 p\left( M_t = m | M_{t-1} = n \right) = \Upsilon_{m,n}, 
\end{equation}
where $m,n \in \{ 1,\ldots, 8\}$. Hence, we can write the full generative model as
\begin{equation}
\label{eq:genmodel}
\begin{split}
p&\left( y_{t:\tau}, S_{t:\tau-1}, M_{t:\tau-1} | y_{\tau-1:0} \right) = \\
&p\left(S_{\tau-1}, M_{\tau-1} \right) \prod_{k = \tau}^{t} p\left( y_k| y_{k-1:k-\tau}, S_{k} \right)\\
&p\left( S_k| S_{k-1}, M_{k}, M_{k-1} \right) p\left( M_k| M_{k-1} \right).
\end{split}
\end{equation}

\subsubsection{Online Bayesian inference} 
Using the above described generative model of the recorded data (Eq.~\ref{eq:genmodel}) we can estimate the posterior movement probability at time step $t$ as
\begin{equation}
\begin{split}
p&\left(M_t| y_{t:0} \right) = \sum_{S_t} p\left(M_t, S_t| y_{t:0} \right) =\\ 
&\sum_{S_t}\frac{p\left(\vec{y}_t| \vec{y}_{t-1:t-p}, S_t, M_t \right)p\left(S_t, M_t| y_{t-1:0}\right)}{p\left(y_t| y_{t-1:0}\right)},
\end{split} 
\end{equation}
where $p\left(S_t, M_t| y_{t-1:0}\right)$ denotes prior distribution over movements $M_t$ and movement segments $S_t$, and $p\left(y_t| y_{t-1:0}\right)$ denotes the normalization constant, which ensures that $\sum_{S_t, M_t} p\left(M_t, S_t| y_{t:0} \right) = 1$. The prior distribution is estimated using the posterior distribution at previous time step and the transition matrices of movements and movement segments (see Eq.~\ref{eq:movtm} and Eq.~\ref{eq:segtm} ), hence
\begin{equation}
\begin{split}
p&\left(S_t, M_t| y_{t-1:0}\right) = \\
\sum_{S_{t-1}, M_{t-1}} &p\left(S_t| S_{t-1}, M_t, M_{t-1} \right) \\
&p\left(M_t| M_{t-1} \right) p\left(S_{t-1}, M_{t-1}|y_{t-1:0}\right).
\end{split}  
\end{equation}
Using the posterior movement probability we define the label $l_t$ of the recorded signal $\vec{y}_t$ at time step $t$ as
\begin{equation}
l_t = \argmax_m p\left(M_t = m| y_{t:0} \right).
\end{equation}

\subsection{Estimation of model parameters}
The accuracy of the above defined classification procedure is fully dependent on the values of the free parameters of the generative model. The parameters that define the VAR models $\{\tau, \mu^{(i,m)}, \Sigma^{(i,m)}, A^{(i,m)}_{1:\tau}\}$ and the transitions between the movement segments ${T^{(m)}_{i,j}, p_{i,m}}$ were estimated using a Viterbi algorithm \cite{logothetis1999expectation,forney1973viterbi} combined with the Expectation Maximisation (EM) \cite{north1998learning}. Importantly, using the Bayesian Information Criterion (BIC) \cite{watanabe2013widely} we have determined that the time lag $\tau$ of the VAR model that provides on average the highest model evidence (highest BIC estimate) for the recorded data corresponds to $\tau = 1$. Hence, all the movements where fitted with the set of five VAR(1) models.

The fitting procedure of the Viterbi EM method goes as follows: (i) sample initial parameter values $\left\{\mu^{(1,m)},\ldots, \mu^{(5,m)},  A^{(1,m)}_{1}, \ldots, A_1^{(5,m)}\right\}$ from a standard multivariate normal distribution and set  $\left\{ \Sigma^{(1,m)}, \ldots, \Sigma^{(5,m)}\right\}$ to identity matrix; (ii) use the Viterbi algorithm to estimate the most likely sequence of segments over the training data for the $m$th movement; (iii) using the most likely sequence of segments estimate the elements of transition matrix of movement segments $T^{(m)}_{i,j}$, and prior segment distribution $p_{i,m}$; (iv) re-estimate the parameters of the VAR models ($\mu^{(1,m)},\ldots,\mu^{(1,m)}$, $A^{(1,m)}_{1}, \ldots, A_1^{(5,m)}$, and $ \Sigma^{(1,m)}, \ldots, \Sigma^{(5,m)}$) using a maximum likelihood estimate, and the most likely segmentation of the training data provided by the Viterbi algorithm; (v) repeat the procedure from step (ii) until the convergence of the log-likelihood $\left(ftol = 10^{-10} \right)$. We have used the implementation of the Viterbi EM algorithm provided by the pyhsmm-autoregressive Python library \cite{johnson2012}.

As the above fitting procedure does not insure convergence to the global optima, we have repeated the estimation procedure for $n = 1000$ different initial conditions (step (i) of the algorithm), and estimated the log-likelihood over the training data set at the end of each repetition. In the end we keep only the parameter values that correspond to the run with highest log-likelihood estimate.

The prior movement probability was set to $p\left(M_0 = 1\right) = 1$, and $p\left(M_0 \neq 1\right) = 0$ as the testing data always started from the resting position. Similarly, we have used the knowledge that the each movement in the testing data set will be either preceded or succeeded by a resting period, to define that movement transition matrix as follows
\begin{equation*}
\Upsilon_{i,j} = \left\{ \begin{array}{cll}
\rho, & \textrm{for }& i = j \\
1-\rho, & \textrm{for }& i=1;\:j\in\{2,\ldots, 8\}\\
\frac{1-\rho}{N-1},& \textrm{for }& i\in\{2,\ldots,8\}; j = 1\\
0, & \textrm{otherwise}
\end{array} \right.
\end{equation*}
where $\rho = 0.999$, and $N = 8$ (total number of different movements across all data sets).

\subsection{Evaluation of the performance of online classifiers}
We have evaluated the effectiveness of the KNN and VAR-HHMM algorithm using a quantitative and a qualitative analysis:
\begin{itemize}
\item For the quantitative analysis we estimated the movement onset detection error and movement classification error of the proposed algorithm in comparison to the KNN algorithm. The EC algorithm served as a benchmark for movement onset detection and movement classification evaluation. The quantitative analysis of classification comprised confusion matrix, precision score and recall score for the VAR-HHMM and the KNN algorithm using the data that was pre-labeled with the offline EC algorithm.

\item For the qualitative analysis we estimated the detection of the movement onset and compared it with the time window derived from SMtSC table. For the movement onset detection we estimated the time difference between the markers labeled by the VAR-HHMM and the KNN algorithms with respect to the EC offline markers. Furthermore, we imposed additional delay constrains to consider possible hardware latencies (Bluetooth, sound card and buffers)
\end{itemize}

To derive the time window from SMtSC data, we have first excluded subjects that had inconsistent responses for more than $50 \%$ of trials. Using this criteria we excluded in total 4 subjects. We then encoded the subject responses (synchronous or asynchronous ratings) as $1$ if they rated the trial as synchronous and $0$ if they rated the trial as asynchronous. This encoding enabled us to perform the summation of individual responses for all trials to estimate rating distribution. Based on this distribution we were able to calculate the range of acceptable sound desynchronization with respect to movement onset.

\section{Results}
\begin{figure*}
\centering
\includegraphics[width=\textwidth]{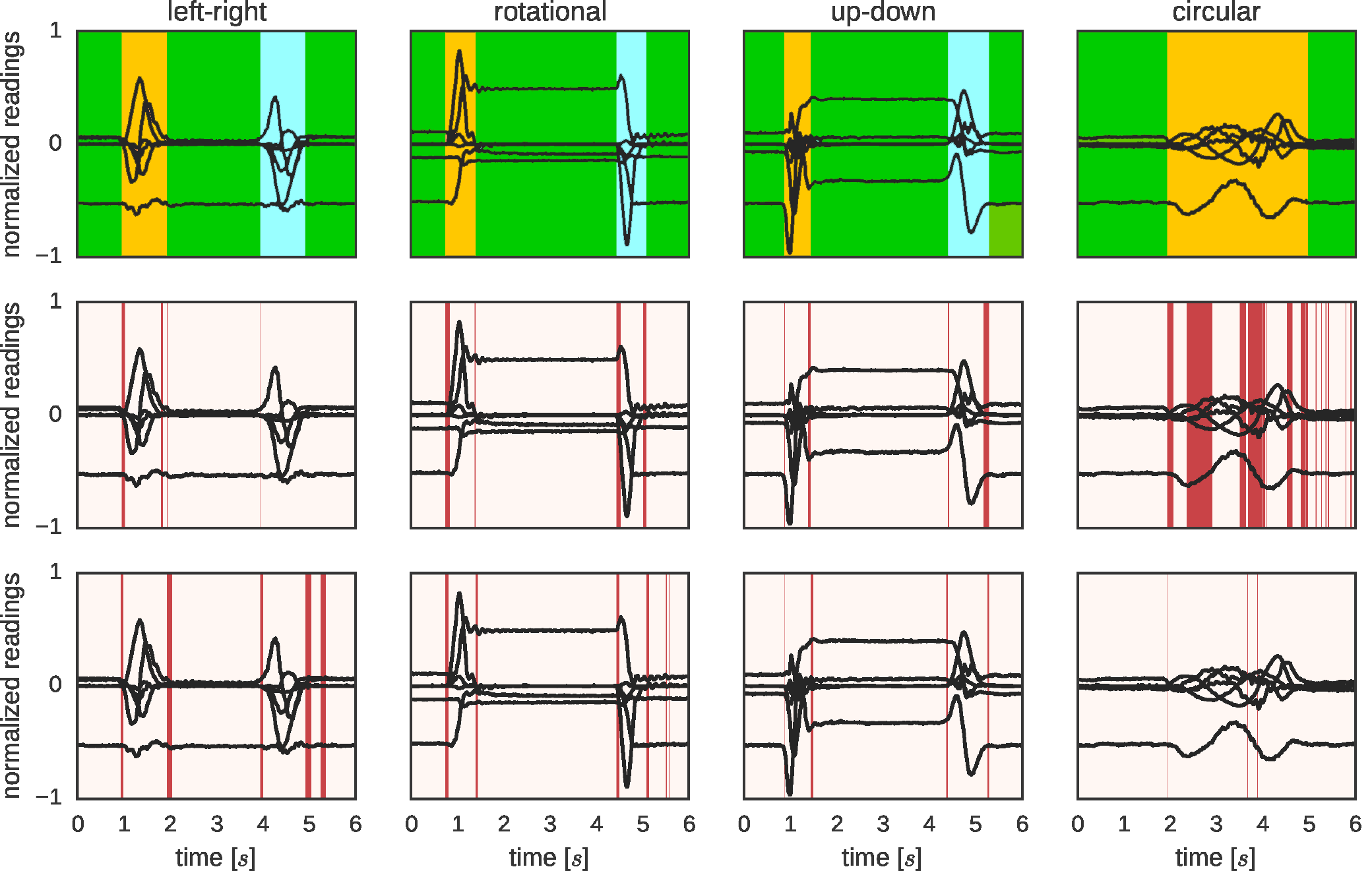}
\caption{A snapshots of the recorded signal from all data sets. (Top row) Movement classified using the offline EC algorithm. (Middle row) Movement classified using the online KNN algorithm. (Bottom row) Movement classified using the online VAR-HHMM algorithm. The red colored regions denote the data points which the corresponding algorithm misclassified with respect to the benchmark classification of the offline EC algorithm.}

\label{fig:rawsignal}
\end{figure*}
\subsection{Quantitative analysis}
We will start the analysis by illustrating the typical difference between the EC, KNN and the VAR-HHMM algorithm. We show in Fig.~\ref{fig:rawsignal} the $6\unit{s}$ long snapshots from four different data sets of recorded movements. The coloured regions in the top row of the graph show the benchmark labels that we determined with the expert classifier. The red shaded areas in the other two rows of the graph show the misclassified data points either by the KNN (middle row) or the VAR-HHMM algorithm (bottom row). One can notice that the biggest difference between the two algorithms is in the case of the circular movements (the right most column); the KNN classifier makes a significant number of misclassifications during the duration of the movement, while VAR-HHMM keeps a proper movement model throughout the snapshot, with only a few misclassified samples during transition between movement segments. Hence, as a next step, we will compare the two algorithms in more details using various classification metrics.

\begin{figure*}
\centering
\subfloat[KNN]{\includegraphics[width=\columnwidth]{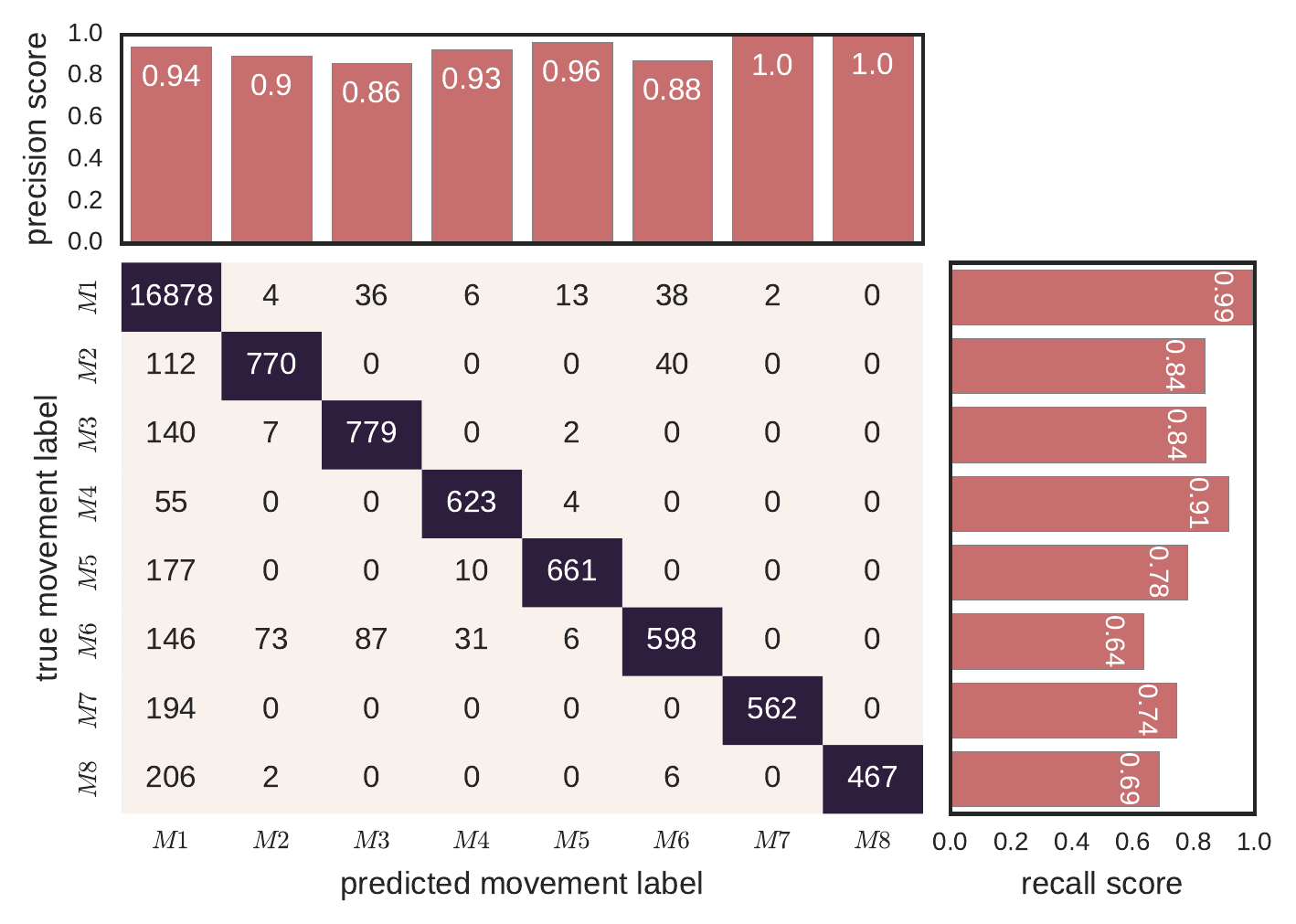}
\label{fig:cmknn}}
\hfil
\subfloat[VAR-HHMM]{\includegraphics[width=\columnwidth]{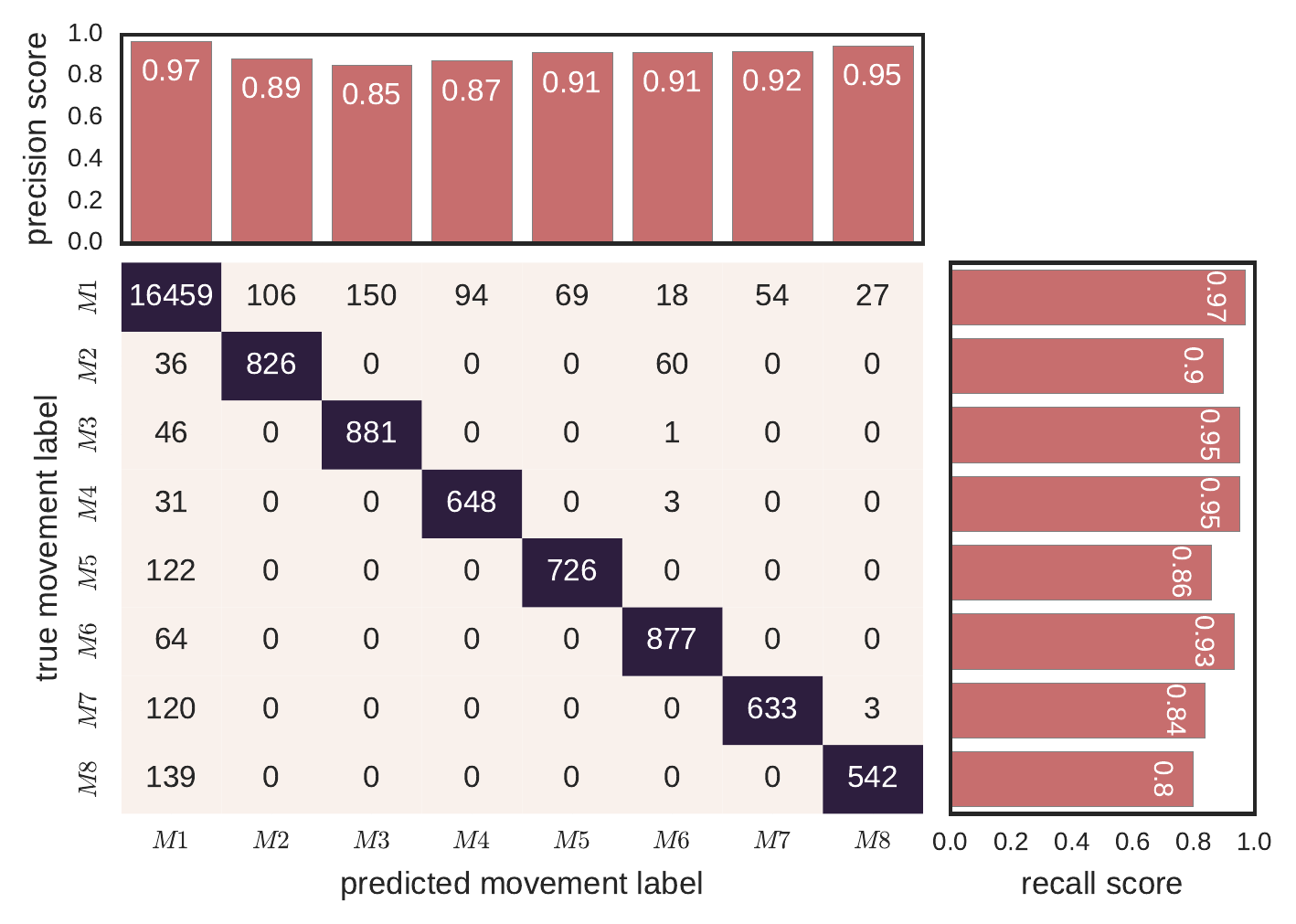}
\label{fig:cmarhhmm}}
\caption{The confusion matrix and the related classification metrics for the two online classifiers.}
\label{fig:cm}
\end{figure*}

To evaluate the classification performance of the two algorithms across all recorded data sets we have estimated the confusion matrices, precision score and recall scores. The overall classification performance of the KNN classifier is shown in Fig.~\ref{fig:cmknn} and of the VAR-HHMM classifier in Fig.~\ref{fig:cmarhhmm}. In ideal conditions, all the time points would be perfectly matched to the labelling obtained with the expert classifier and one would observe nonzero values only on the diagonal of the confusion matrix. However, due to the intrinsic variability of human movements, in practice, we find that certain amount of data samples is misclassified (non-zero off-diagonal elements in the two confusion matrices). 

To estimate the average, per movement, classification accuracy we have estimated the precision score and the recall scores, which are shown on the top and right graphs, respectively, of Fig.~\ref{fig:cmknn} and Fig.~\ref{fig:cmarhhmm}. The precision score measures the fraction of data points that were correctly classified with respect to all the samples that are assigned to the particular movement. Similarly, the recall score measures the fraction of data points that were correctly classified with respect to all the data points that truly belong to the specific movement label (as defined with the expert classifier). From Fig.~\ref{fig:cmknn} we can see that the most of the KNN classification errors come from classifying a movement ($M2, \ldots, M8$) as a resting state $M1$. Also, the circular movement is often misclassified as a left-right or an up-down movement as it incorporates fragments of those movements during its execution. In contrast, the VAR-HHMM classifier significantly improves $M2-M8$ recall scores (it correctly assigns a data point to its true movement label; see Fig.~\ref{fig:cmarhhmm}) with a slight decrease in the precision scores due to a more frequent labelling of the rest samples as movement samples.

\begin{figure}[!ht]
\centering
\includegraphics[width=1\columnwidth]{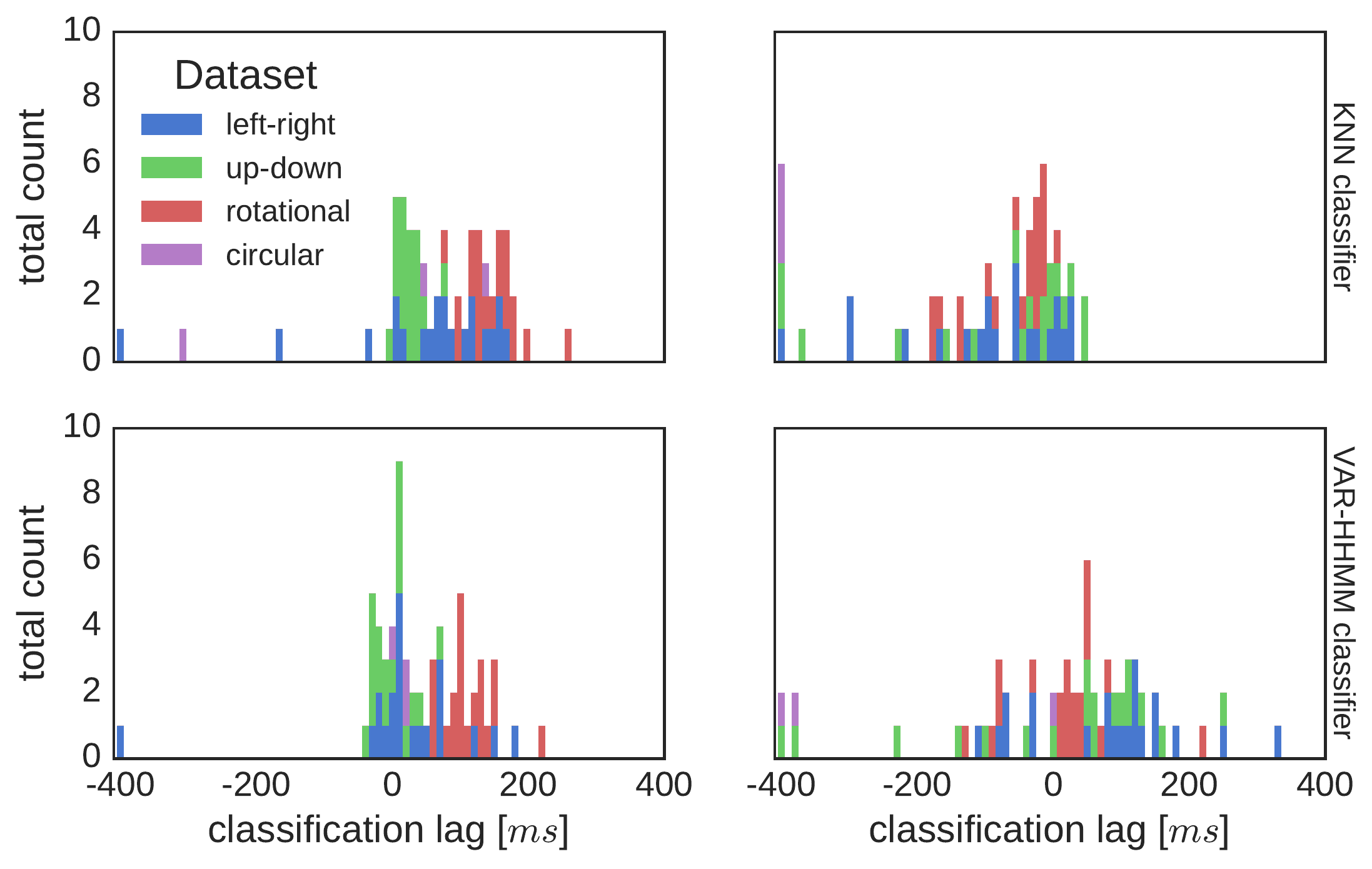}
\caption{ Histograms of classification lags. (Top row) KNN classifier, (bottom row) VAR-HHMM classifier. (Left column) Movement onset, (right column) movement end. Detections are labeled by different colors. It can be noted that both algorithms classify up-down movements onset early and reliably. Left-right movement onsets present much easier task for VAR-HHMM to handle compared to the KNN, resulting in less outliers, smaller variability and earlier detections.  The most significant difference between algorithms is in the case of circular movement onset which is robustly detected by the VAR-HHMM but very poorly by the KNN.}
\label{fig:lags}
\end{figure}

To better understand the causes of the mixing movement and resting related data samples we have looked at the properties of movement onset and movement endpoint detection. In Fig.~\ref{fig:lags} we show the histograms of the classification lags (the time separation between the detected and true start and end points of the movements). From the movement onset histogram, we can infer that some of the misclassified resting samples can be attributed to the too early the detection of the movement onset and early detection of the movement end-points. Notice that the VAR-HHMM classifier registers movement onset on average earlier than the KNN classifier. Similarly, the KNN classifier registers the movement endpoint on average earlier than the VAR-HHMM classifier. This early detection of movement onset and endpoint explains why VAR-HHMM classifier shows more resting samples classified as movement and why KNN classifier shows more movement samples labelled as rest samples.

\begin{figure}[!ht]
\centering
\includegraphics[width=\columnwidth]{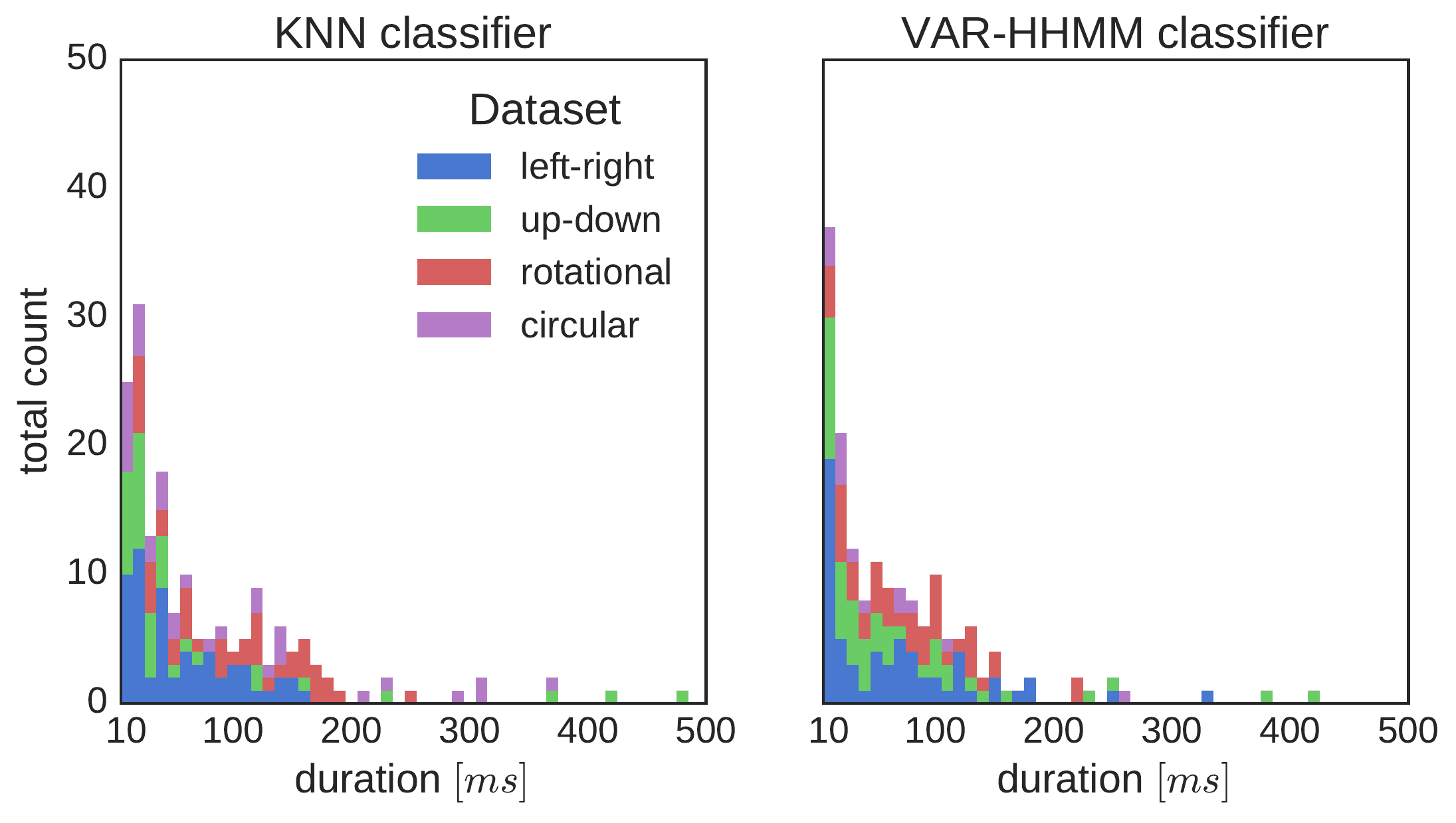}
\caption{ Histograms of durations of misclassified segments for the two types of classifiers. Main 4 (first 4) components of the KNN histogram contain $14 \%$, $18 \%$, $7.5 \%$, $10.5 \%$ percentages of total errors in classification. These percentages for the VAR-HHMM are: $22 \%$, $12.5 \%$, $7 \%$, $5 \%$. Although number of misclassified segments are similar for both algorithms (174 for KNN and 166 for VAR-HHMM) total number of misclassified samples is significantly higher ($22 \%$) for KNN due to greater dispersion of segments durations.}
\label{fig:durs}
\end{figure}

Finally, to compare the stability (the ability to maintain the active movement model in the presence of perturbations) and the flexibility (the transition to the actually active movement model without inherent time lag) of the two algorithms we have estimated the duration of misclassified blocks of samples. The histogram of the durations of misclassified samples is shown in Fig.~\ref{fig:durs}. For sound orientated detections, the worst case scenario is switching to a false model and keeping it for a long time. Importantly, in both algorithms, the duration of the falsely labelled blocks of samples exponentially decline. Although it appears that the VAR-HHMM algorithm has on average shorter duration of misclassified blocks, the difference is subtle (approx. $10\unit{ms}$) and non-significant as we cannot reject the null hypothesis ($p > 0.05$, the two tailed F-test).

\subsection{Qualitative analysis}
\begin{figure}[!ht]
\centering
\includegraphics[width=\columnwidth]{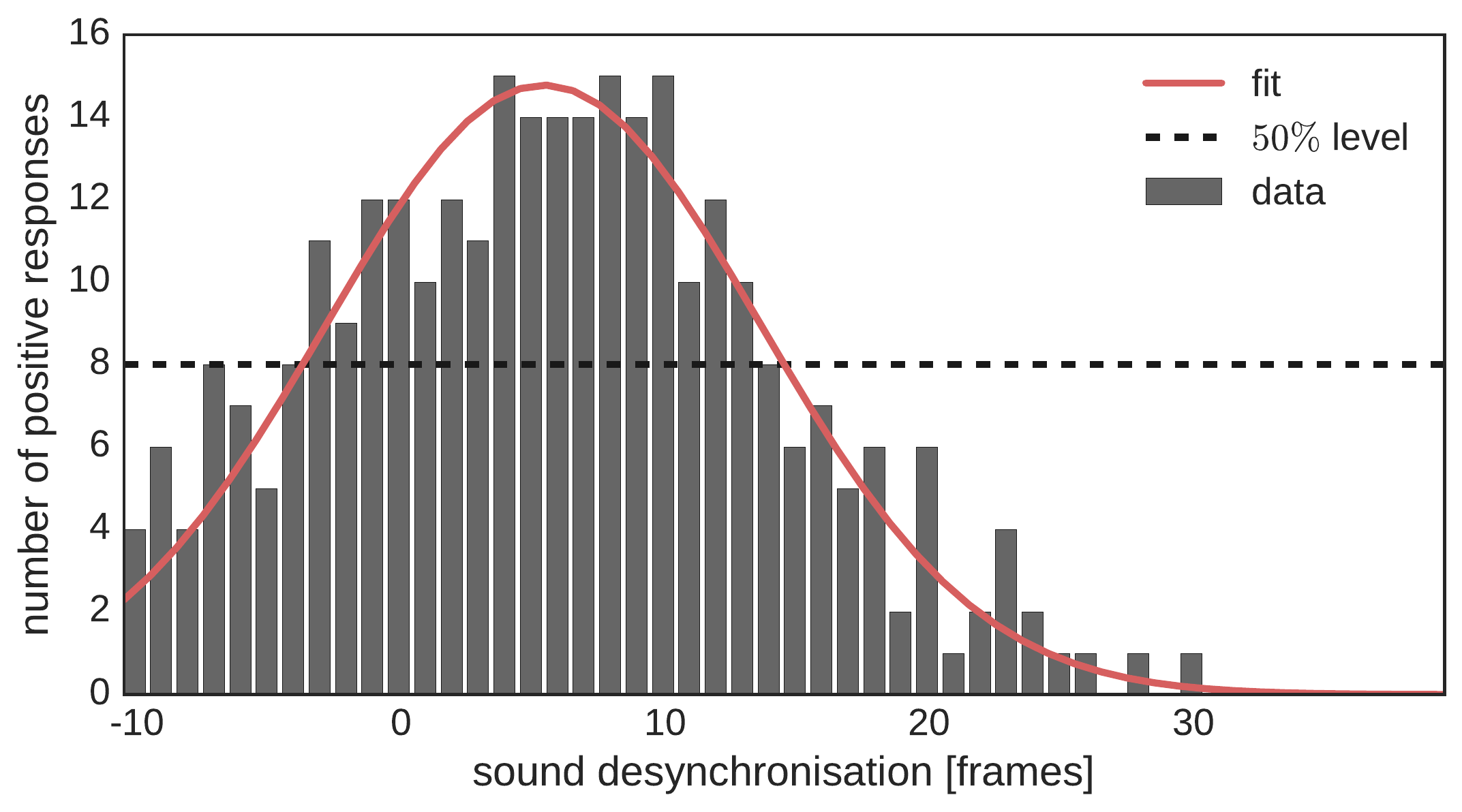}
\caption{Histogram of the total number of positive responses with respect to the sound desynchronisation. Blue bars dots denote the total number of times that the participants rated positively the relation between the movement and the lagged sound for a given level of desynchronization. The red line shows the fitted normal density ($\mu = 5.88$ frames, $\sigma = 8.23$ frames) multiplied by the total number of positive responses $N = 305$. The black dashed line shows the level corresponding to half of the total number of participants.%
}
\label{fig:data}
\end{figure}

Although the classification metric allows us to compare the performance of the two classifiers directly it tells us little about their performance in real world applications. In other words, is the classification good enough to be useful in the Movezik system for movement to sound mapping? To test this, we used the experimental data to define the bounds on acceptable real-time classification accuracy. In Fig.~\ref{fig:data} we show the histogram of participants' positive responses (rating movement-sound relationship as synchronised) for various desynchronisation levels. As the Lilliefors' test cannot reject the null hypothesis that the data follow a normal distribution with the significance level $p=0.001$, we have defined the acceptable desynchronisation between movement and sound onset as a range that spans one standard deviation away from the mean ($68\%$ probability mass interval). Interestingly, this range also matches the range of all lags which more than half of the participants rated positively. Hence, the acceptable movement onset delay spans the interval from -3 frames ($-48\unit{ms}$) up to 13 frames ($208\unit{ms}$). Importantly, these values are very close to what others have reported in similar audio-visual synchronisation experiments \cite{vatakis2006audiovisual,zampini2005audio}.

\begin{figure}[!ht]
\centering
\includegraphics[width=\columnwidth]{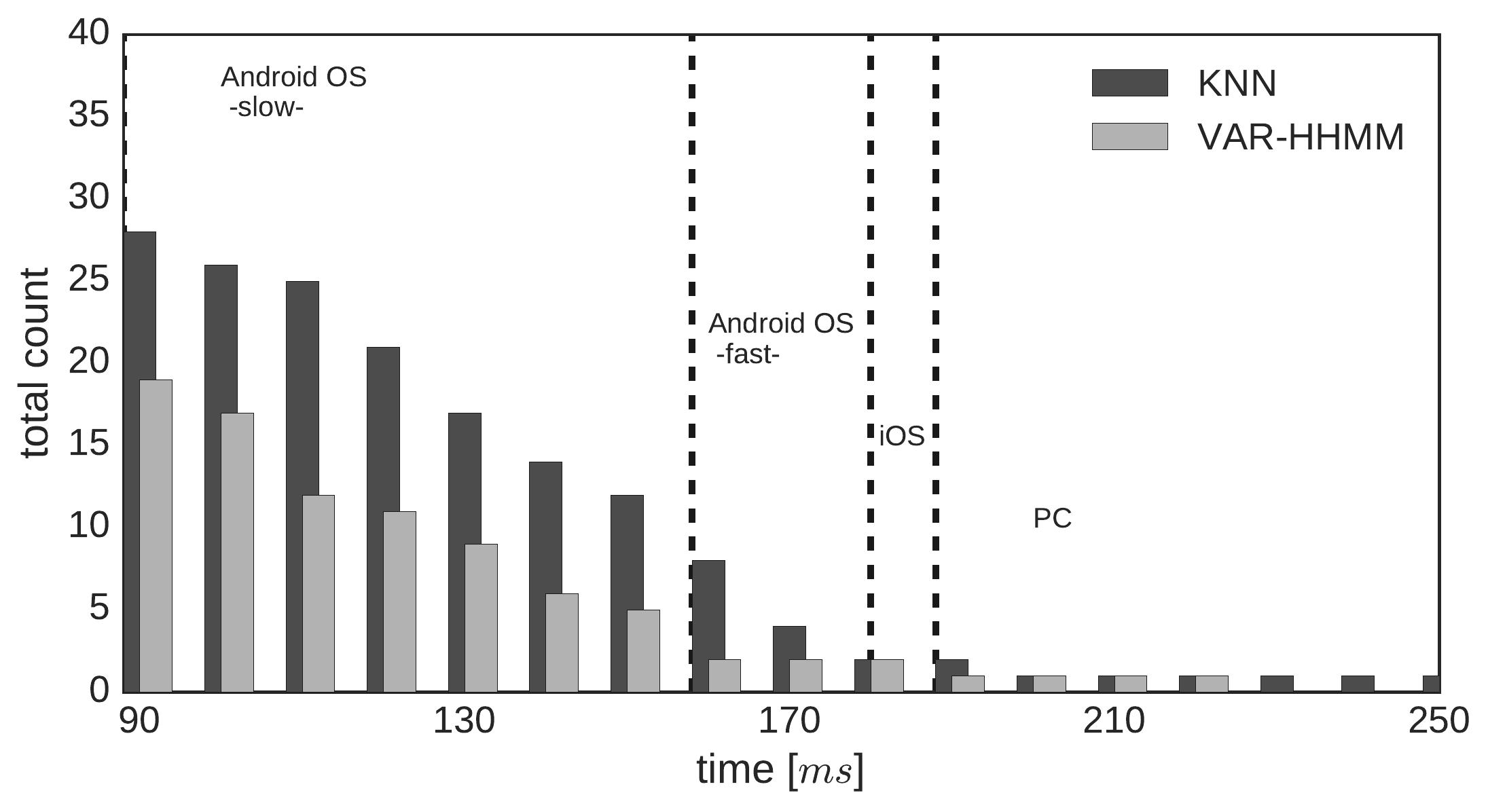}
\caption{The total number of movements that were classified correctly only after the given time threshold. The vertical dashed lines denote the lag threshold above which the movements and sounds would be perceivably desynchronised on a specific hardware-software platform.}
\label{fig:cumerr}
\end{figure}

In addition to the experimentally derived latency threshold, we have also considered hardware and software limitations of the state-of-the-art devices and platforms. While Bluetooth latencies across different devices are unified, ranging from $20\unit{ms}$ to $40\unit{ms}$, latencies originating from sound card responses fall in a range from less than $1\unit{ms}$ in the case of PC, up to $100\unit{ms}$ in some Android devices \cite{superpower} (Android latency issue). With this in mind, we reduced acceptable detection lag range for different hardware and software platforms up to $88\unit{ms}$ in the case of Android devices with high sound latencies (Fig.~\ref{fig:cumerr}). We find that both algorithms had only one (out of 126 movements) perceivably delayed movement onset detection for the PC implementation, which corresponds to zero hardware latency. For the iOS based implementation the acceptable delay is reduced by $8\unit{ms}$ which leads to two (out of 126) perceivably delayed detections for both algorithms. The performance of two algorithm starts to diverge only after decreasing the range of acceptable latencies to include Android based implementations. The VAR-HHMM classifier outperformed the KNN classifier for both fast Android devices ($8$ delayed detections by KNN vs $2$ delayed detections by VAR-HHMM) and slow Android devices ($28$ delayed detections by KNN vs $19$ delayed detections by VAR-HHMM). This results imply that further improvements to the VAR-HHMM classifier will be required for applying the proposed Movezik system in slow Andorid devices.

\section{Discussion}

Here we presented a new sensorised hardware platform coupled with a novel methodology for learning dynamical models of movement trajectories. This methodology is suitable for fast movement onset detection and movement classification. The custom made hardware (Movezik) is based on a wireless IMU sensor suitable for human motion tracking, while the key control algorithms are platform independent. The proposed movement classification algorithm is based on a vector autorgressive processes (VAR) coupled with hierarchical hidden Markov models (HHMM). 

As the main goal of the presented system is to interconnect human movements with sounds, the movement detection algorithm should provide fast and reliable detection of movements that can be translated into sound in real-time in order to create a clear perceivable correlation between a movement and a sound. To evaluate the qualitative and quantitative performance of the algorithm we used two benchmarks, the k-nearest neighbors (KNN) classifier and an offline expert classifier (EC). 

The basic quantitative evaluation criterion was movement classification error which was calculated in respect to offline EC and benchmarked with common KNN method. The test was implemented on the recorded sequence comprising $126$ arm movements that are grouped on $7$ distinct movements, plus segments of resting arm positions between the movements. Overall, both VAR-HHMM and KNN algorithms produce similar classification outcomes, with KNN exhibiting on average higher precision scores and VAR-HHMM outperforming KNN in class recall for all but one movement (compare Fig.~\ref{fig:cmknn} and Fig.~\ref{fig:cmarhhmm}). These differences between the two online classifiers can be explained using the fact that VAR-HHMM detect movement onsets slightly before the EC standard, whereas that KNN classifiers detects end of the movements before the EC standard (see Fig.~\ref{fig:lags}). In addition, both algorithms produce similar distribution of the duration of misclassified blocks of data (see Fig.~\ref{fig:durs}). As the the majority of misclassified blocks of data lasts only a couple of samples, it is possible to impose additional switching criterion to make compromise between the movement onset detections latency and the misclassification count. 

Importantly, we find several advantages of VAR-HHMM algorithm compared to the KNN algorithm:
\begin{itemize}
\item VAR-HHMM more accurately classifies complex movement trajectories (circular movement of the arm; see Fig.~\ref{fig:cm}) that can be decomposed on simpler movement modules. While KNN assigns parts of the complex movement to similar simpler movements, VAR-HHMM detects and holds correct model state throughout circular movement (with only occasional model switching; see Fig.~\ref{fig:rawsignal}). This VAR-HHMM performance feature makes it advantageous in applications with large number of different movements, including complex limb trajectories.

\item VAR-HHMM exhibits less outliers during the detection of movement onset and movement termination (see Fig.~\ref{fig:lags}). We consider this to be hard errors as they appear completely uncorrelated with the movement transitions.

\item VAR-HHMM exhibits smaller dispersion of the of the movement onset around the EC standard compared to the KNN algorithm. This difference is pronounce when narrowing acceptable latencies ranges that results in steeper increase of detection errors by the KNN.

\item VAR-HHMM shows potentially better performance then the offline EC algorithm as the fraction of movement onsets are detected before (see range $[-50, 0]\unit{ms}$ in Fig.~\ref{fig:lags}) the time stamp provided by the EC algorithm.
 
\end{itemize}

These advantages of the VAR-HHMM algorithm are also reflected in the qualitative analysis of the performance of the two algorithms. The qualitative analysis was based on the subjective movement-to-sound criterion (SMtSC) that was obtained from a behavioral experiment in which we asked participants to rate synchronisation levels between movements and sounds. The SMtSC measure suggests that the experimental trials with onset lag smaller than $208\unit{ms}$ are perceived as synchronous. Using these threshold of subjectively acceptable lags for the detection of movement onset we find that both algorithms fulfill detect movement onset before this threshold for more than $99 \%$ of movements (each had one detection outside this range). Furthermore, when accounting for an additive lag originating from the common wireless protocols and latencies related to executions of instructions between the application and sound card driver (processing lag due to the influence of background processes of a non-real-time operating system was omitted) we find that the percentage of detections above the perceptual threshold remained lower in the case of VAR-HHMM algorithm compared to the KNN algorithm (see Fig.~\ref{fig:cumerr}).

Beside the above discussed advantages in the classification performance, another practical advantage of the VAR-HHMM based algorithm to train additional movement models without full recalculation of a models data base. This feature makes it more suitable in dynamic interfaces intended for adaptable human-computer interactions with increasing number of human inputs. In addition, a system that is based on the VAR-HHMM algorithm could potentially adjust models' parameters in real-time. This is clear advantage over KNN and similar implementations that require substantially more processing time to recalculate models. In addition, KNN algorithm execution involves relatively large memory (storing whole training set for time based signal can easily overshoot 1 MB) and processor cycles per obtained sample (KNN needs to calculate a metric for all the samples of the training set) while VAR-HHMM algorithm needs only a fraction of a memory for storing linear models and considerably less processing power. This advantage permits execution of the VAR-HHMM algorithm in real-time on the vast majority of microprocessor architectures, even low-power devices that can be worn on the body during prolonged use.

Finally, in spite of several advantages of the VAR-HHMM algorithm over established standards such as KNN classifier, the improvements are still insufficient for immediate real world applications. The number of false positives and negatives are still too large for applications that require a robust control paradigm, and the lag for detection of movement onset is still large for the majority of the mid to low performance mobile devices that are currently available. In future studies we hope to improve the classification algorithm by testing out extensions of the linear VAR models to the nonlinear domain with nonlinear autoregressive neuronal networks \cite{billings2013} or convolutional neuronal networks \cite{Taylor2010}. We expect that the improving dynamical models of movement trajectories will bring further improvements to the classification performance.   

\section{Conclusion}
Besides the application of the presented system as a movement-to-sound interface the presented methodology and technology can be easily extended and applied to any body-to-machine system. For this to be possible, the classification scheme should eventually achieve fast and reliable detection of movements that can be translated into an arbitrary sequence of commands in real-time and create a perceivable relation between the cause, a movement, and the effect, a command.

In conclusion, the here proposed system opens exciting possibilities for novel forms of creative expression and a multitude of still unexplored applications.

\section*{Acknowledgment}
This work was partly supported by the Ministry of Education, Science and Technological Development of the Republic of Serbia (Project No. OI175016).  

\bibliographystyle{IEEEtran}
\bibliography{biblio}

\begin{thebibliography}{10}
\providecommand{\url}[1]{#1}
\csname url@samestyle\endcsname
\providecommand{\newblock}{\relax}
\providecommand{\bibinfo}[2]{#2}
\providecommand{\BIBentrySTDinterwordspacing}{\spaceskip=0pt\relax}
\providecommand{\BIBentryALTinterwordstretchfactor}{4}
\providecommand{\BIBentryALTinterwordspacing}{\spaceskip=\fontdimen2\font plus
\BIBentryALTinterwordstretchfactor\fontdimen3\font minus
  \fontdimen4\font\relax}
\providecommand{\BIBforeignlanguage}[2]{{%
\expandafter\ifx\csname l@#1\endcsname\relax
\typeout{** WARNING: IEEEtran.bst: No hyphenation pattern has been}%
\typeout{** loaded for the language `#1'. Using the pattern for}%
\typeout{** the default language instead.}%
\else
\language=\csname l@#1\endcsname
\fi
#2}}
\providecommand{\BIBdecl}{\relax}
\BIBdecl

\bibitem{jensenius2007action}
A.~R. Jensenius, ``Action-sound: Developing methods and tools to study
  music-related body movement,'' Ph.D. dissertation, Department of Musicology,
  University of Oslo, 2007.

\bibitem{gritten2006music}
A.~Gritten and E.~King, \emph{Music and gesture}.\hskip 1em plus 0.5em minus
  0.4em\relax Ashgate Publishing, Ltd., 2006.

\bibitem{cadoz2000gesture}
C.~Cadoz and M.~M. Wanderley, ``Gesture-music,'' \emph{Trends in gestural
  control of music}, pp. 71--93, 2000.

\bibitem{bongers2000physical}
B.~Bongers, ``Physical interfaces in the electronic arts,'' \emph{Trends in
  gestural control of music}, pp. 41--70, 2000.

\bibitem{paradiso1997electronic}
J.~A. Paradiso, ``Electronic music: new ways to play,'' \emph{IEEE spectrum},
  vol.~34, no.~12, pp. 18--30, 1997.

\bibitem{ferreira2008sound}
P.~P. Ferreira, ``When sound meets movement: Performance in electronic dance
  music,'' \emph{Leonardo Music Journal}, vol.~18, pp. 17--20, 2008.

\bibitem{stuart2003object}
C.~Stuart, ``The object of performance: Aural performativity in contemporary
  laptop music,'' \emph{Contemporary Music Review}, vol.~22, no.~4, pp. 59--65,
  2003.

\bibitem{schloss2003using}
W.~A. Schloss, ``Using contemporary technology in live performance: The dilemma
  of the performer,'' \emph{Journal of New Music Research}, vol.~32, no.~3, pp.
  239--242, 2003.

\bibitem{tanaka2000musical}
A.~Tanaka, ``Musical performance practice on sensor-based instruments,''
  \emph{Trends in Gestural Control of Music}, vol.~13, no. 389-405, p. 284,
  2000.

\bibitem{winkler1995making}
T.~Winkler, ``Making motion musical: Gesture mapping strategies for interactive
  computer music,'' in \emph{ICMC Proceedings}, 1995, pp. 261--264.

\bibitem{mitchell2011soundgrasp}
T.~J. Mitchell, ``Soundgrasp: A gestural interface for the performance of live
  music,'' in \emph{In: International Conference on New Interfaces for Musical
  Expression (NIME), Oslo, Norway, 30 May - 1 June 2011}, 2011.

\bibitem{wang2008chuck}
G.~Wang, \emph{The chuck audio programming language. a strongly-timed and
  on-the-fly environ/mentality}.\hskip 1em plus 0.5em minus 0.4em\relax
  Princeton University, 2008.

\bibitem{bevilacqua2007wireless}
F.~Bevilacqua, F.~Gu{\'e}dy, N.~Schnell, E.~Fl{\'e}ty, and N.~Leroy, ``Wireless
  sensor interface and gesture-follower for music pedagogy,'' in
  \emph{Proceedings of the 7th international conference on New interfaces for
  musical expression}.\hskip 1em plus 0.5em minus 0.4em\relax ACM, 2007, pp.
  124--129.

\bibitem{iazzetta2000meaning}
F.~Iazzetta, ``Meaning in musical gesture,'' 2000.

\bibitem{choi1998motion}
I.~Choi, ``From motion to emotion: Synthesis of interactivity with gestural
  primitives,'' \emph{Emotional and Intelligent: The Tangled Knot of
  Cognition}, pp. 22--25, 1998.

\bibitem{goldstein1998gestural}
M.~Goldstein, ``Gestural coherence and musical interaction design,'' in
  \emph{Systems, Man, and Cybernetics, 1998. 1998 IEEE International Conference
  on}, vol.~2.\hskip 1em plus 0.5em minus 0.4em\relax IEEE, 1998, pp.
  1076--1079.

\bibitem{bahn2001physicality}
C.~Bahn, T.~Hahn, and D.~Trueman, ``Physicality and feedback: a focus on the
  body in the performance of electronic music,'' in \emph{Proceedings of the
  International Computer Music Conference}, vol. 2001, 2001, pp. 44--51.

\bibitem{ephraim2002hidden}
Y.~Ephraim and N.~Merhav, ``Hidden markov processes,'' \emph{IEEE Transactions
  on information theory}, vol.~48, no.~6, pp. 1518--1569, 2002.

\bibitem{yang2000some}
M.~Yang, ``Some properties of vector autoregressive processes with
  markov-switching coefficients,'' \emph{Econometric Theory}, vol.~16, no.~01,
  pp. 23--43, 2000.

\bibitem{north1998learning}
B.~North and A.~Blake, ``Learning dynamical models using
  expectation-maximisation,'' in \emph{Computer Vision, 1998. Sixth
  International Conference on}.\hskip 1em plus 0.5em minus 0.4em\relax IEEE,
  1998, pp. 384--389.

\bibitem{logothetis1999expectation}
A.~Logothetis and V.~Krishnamurthy, ``Expectation maximization algorithms for
  map estimation of jump markov linear systems,'' \emph{IEEE Transactions on
  Signal Processing}, vol.~47, no.~8, pp. 2139--2156, 1999.

\bibitem{forney1973viterbi}
G.~D. Forney, ``The viterbi algorithm,'' \emph{Proceedings of the IEEE},
  vol.~61, no.~3, pp. 268--278, 1973.

\bibitem{bishop1995neural}
C.~M. Bishop, \emph{Neural networks for pattern recognition}.\hskip 1em plus
  0.5em minus 0.4em\relax Oxford university press, 1995.

\bibitem{cacoullos2014discriminant}
T.~Cacoullos, \emph{Discriminant analysis and applications}.\hskip 1em plus
  0.5em minus 0.4em\relax Academic Press, 2014.

\bibitem{lachenbruch1979discriminant}
P.~A. Lachenbruch and M.~Goldstein, ``Discriminant analysis,''
  \emph{Biometrics}, pp. 69--85, 1979.

\bibitem{meyer2015support}
D.~Meyer and F.~T. Wien, ``Support vector machines,'' \emph{The Interface to
  libsvm in package e1071}, 2015.

\bibitem{du2014support}
K.-L. Du and M.~Swamy, ``Support vector machines,'' in \emph{Neural Networks
  and Statistical Learning}.\hskip 1em plus 0.5em minus 0.4em\relax Springer,
  2014, pp. 469--524.

\bibitem{davis2006relationship}
J.~Davis and M.~Goadrich, ``The relationship between precision-recall and roc
  curves,'' in \emph{Proceedings of the 23rd international conference on
  Machine learning}.\hskip 1em plus 0.5em minus 0.4em\relax ACM, 2006, pp.
  233--240.

\bibitem{fawcett2006introduction}
T.~Fawcett, ``An introduction to roc analysis,'' \emph{Pattern recognition
  letters}, vol.~27, no.~8, pp. 861--874, 2006.

\bibitem{haptic}
V.~Hayward, O.~R. Astley, M.~Cruz-Hernandez, D.~Grant, and
  G.~Robles-De-La-Torre, ``Haptic interfaces and devices,'' \emph{Sensor
  Review}, vol.~24, no.~1, pp. 16--29, 2004.

\bibitem{vatakis2006audiovisual}
A.~Vatakis and C.~Spence, ``Audiovisual synchrony perception for music, speech,
  and object actions,'' \emph{Brain research}, vol. 1111, no.~1, pp. 134--142,
  2006.

\bibitem{zampini2005audio}
M.~Zampini, S.~Guest, D.~I. Shore, and C.~Spence, ``Audio-visual simultaneity
  judgments,'' \emph{Perception \& psychophysics}, vol.~67, no.~3, pp.
  531--544, 2005.

\bibitem{peirce2007psychopy}
J.~W. Peirce, ``Psychopy—psychophysics software in python,'' \emph{Journal of
  neuroscience methods}, vol. 162, no.~1, pp. 8--13, 2007.

\bibitem{hechenbichler2004weighted}
K.~Hechenbichler and K.~Schliep, ``Weighted k-nearest-neighbor techniques and
  ordinal classification,'' in \emph{Discussion Paper 399, SFB 386}, 2006.

\bibitem{scikit-learn}
F.~Pedregosa, G.~Varoquaux, A.~Gramfort, V.~Michel, B.~Thirion, O.~Grisel,
  M.~Blondel, P.~Prettenhofer, R.~Weiss, V.~Dubourg, J.~Vanderplas, A.~Passos,
  D.~Cournapeau, M.~Brucher, M.~Perrot, and E.~Duchesnay, ``Scikit-learn:
  Machine learning in {P}ython,'' \emph{Journal of Machine Learning Research},
  vol.~12, pp. 2825--2830, 2011.

\bibitem{watanabe2013widely}
S.~Watanabe, ``A widely applicable bayesian information criterion,''
  \emph{Journal of Machine Learning Research}, vol.~14, no. Mar, pp. 867--897,
  2013.

\bibitem{johnson2012}
M.~Johnson, ``pyhsmm-autoregresive,''
  https://github.com/mattjj/pyhsmm-autoregressive, 2012.

\bibitem{superpower}
\BIBentryALTinterwordspacing
Superpowered. (2016) Audio technology and audio apis optimized for mobile
  processor architectures. Accessed: 2016-08-08. [Online]. Available:
  \url{http://superpowered.com/latency}
\BIBentrySTDinterwordspacing

\bibitem{billings2013}
S.~A. Billings, \emph{Nonlinear system identification: NARMAX methods in the
  time, frequency, and spatio-temporal domains}.\hskip 1em plus 0.5em minus
  0.4em\relax John Wiley \& Sons, 2013.

\bibitem{Taylor2010}
\BIBentryALTinterwordspacing
G.~W. Taylor, R.~Fergus, Y.~LeCun, and C.~Bregler, ``Convolutional learning of
  spatio-temporal features,'' in \emph{Proceedings of the 11th European
  Conference on Computer Vision: Part VI}, ser. ECCV'10.\hskip 1em plus 0.5em
  minus 0.4em\relax Berlin, Heidelberg: Springer-Verlag, 2010, pp. 140--153.
  [Online]. Available: \url{http://dl.acm.org/citation.cfm?id=1888212.1888225}
\BIBentrySTDinterwordspacing

\end{thebibliography}

\end{document}